\documentclass[reprint, 11pt,
onecolumn,notitlepage, tightenlines, longbibliography]{revtex4-1}

\usepackage{graphicx} 
\graphicspath{{./figures/}}
\usepackage{float}
\usepackage{array} 
\usepackage{paralist} 
\usepackage{subfig}
\usepackage{fancyhdr} 
\usepackage[sort&compress]{natbib}
\usepackage{mathtools}

\usepackage{amsmath,amsfonts,amsthm}

\newtheorem{as}{Assumption}{\bfseries}{\itshape}
\newtheorem{rem}{Remark}{\bfseries}{\itshape}
{\bfseries}{\itshape}

\newcommand{\ket}[1]{\ensuremath{\left|#1\right\rangle}} 
\newcommand{\bra}[1]{\ensuremath{\left\langle#1\right|}} 

\renewcommand{\bf}[1]{\ensuremath{\mathbf{#1}}}

\begin{document}

\title{A Quantum Approach to Subset-Sum and Similar Problems}

\author{Ammar~Daskin}
\affiliation{Department of Computer Engineering, Istanbul Medeniyet University, Kadikoy, Istanbul, Turkey
	\email{adaskin25@gmail.com}} 

\date{Received: date / Accepted: date}
	\begin{abstract}
		In this paper, we study the subset-sum problem  by using  a quantum heuristic approach similar to the verification circuit of quantum Arthur-Merlin games \cite{marriott2005quantum}. 
		Under described certain assumptions, we show that the exact solution of the subset sum problem my be obtained in polynomial time and  the exponential speed-up over the classical algorithms may be possible. 
		We give a numerical example and discuss the complexity of the approach and its further application to the  knapsack problem.
	\end{abstract}
\maketitle
	
	\section{Introduction}
	Subset-sum \cite{kellerer2004introduction} is a widely-studied NP-complete problem formally expressed as follows:
	Given a set of integer elements $V = \{v_1, \dots, v_n\}$ and a target value $W$, determine if there is a subset, $S$, of $V$ whose sum is equal to $W$. 
	In the associated optimization problem, the subset $S$ with the maximum sum \underline{less than}  $W$ is searched. 
	The exact solution for this problem can be found by first computing the sum of elements for each possible $S$ and then selecting the maximum among those whose sum is less than $W$. Clearly, this algorithm would take exponential-time in the number of elements. Another means to solve this problem is through dynamic programming which requires $O(nW)$ time.  
	This is also exponential in the required number of bits to represent $W$: If $W=2^m$ and $m \approx n$, then the running time is $O(n2^m) =O(n2^n) $. There are also many forms of polynomial time approximation algorithms applied to the subset problem. For an overall review of subset-sum problems and the different algorithms, we recommend the book by Keller et al. \cite{kellerer2004introduction}.

	Quantum algorithms in general provides computational speed-up over the classical counter parts. 
	Quantum walk algorithm presented for element distinctness \cite{ambainis2007quantum} is applied to the subset problems \cite{childs2005}. The computational complexity of this algorithm is shown to be bounded by $O(n^{|L|})$, where $n$ is the number of items and $|L|$ is the subset size. 
	In quantum computing, the cases where exponential speedups are possible are generally related to hidden subgroup problems: a few examples of these cases are the factoring \cite{shor1994algorithms}, the dihedral hidden subgroup problem \cite{regev2004quantum} and some lattice problems \cite{kuperberg2005subexponential}. A review of the algorithms giving the exponential speedups in the solutions of algebraic problems are given in Ref.\cite{childs2010quantum}. There are also quantum optimization algorithms such as the ones in the adiabatic quantum computation \cite{farhi2000quantum,farhi2001quantum,aharonov2008adiabatic} applied to different NP problems \cite{lucas2014ising} and the quantum approximate optimization algorithm \cite{farhi2014quantum}  applied to the NP-hard problems. For a further review on general quantum algorithms, please refer to Ref. \cite{bacon2010recent,mosca2012quantum,montanaro2015quantum}, or to the introductory books \cite{nielsen2002quantum,kaye2007introduction}.  
	
	It is known that having the ability of a post-selected quantum computing (an imaginary computing model), one can obtain the result of a Grover search problem in $O(1)$  \cite{farhi2016quantum,aaronson2005quantum}. 
	This ability would also lead a quantum computer to solve NP-complete problems in polynomial time.  
	Although this model is imaginary, it still provides an insight to see one of the differences between quantum and classical computers on the solution of NP-complete problems: 
	i.e., mainly a quantum computer can generate all the solution space and mark the correct answer in polynomial time, which is not possible on classical computers. 
	However, this information, the marked item, can only be obtained by an observer with  an exponential overhead (which makes the computational complexity exponential in the number of qubits.).
	This motivates to do research on the applications of the algorithms  such as Grover's search algorithm \cite{grover1997quantum} to the special cases of NP-complete problems so as to  gain at least some speed-up over the classical algorithms.

	The Grover search algorithm  in quantum computing provides quadratic computational-speedup over the classical brute force search algorithm. 
	It is well-known that the employment of this algorithm in general yields a quadratic speed-up also in the  exact solutions of NP-complete problems (please see the explanation for Hamiltonian cycle given in page 263 of Ref.\cite{nielsen2002quantum}). 
The algorithm also plays important role in quantum Arthur-Merlin games \cite{kitaev1997quantum,marriott2005quantum} and applied along with the phase estimation algorithm to NP-complete problems such as 3-SAT and k-local Hamiltonian problems\cite{kitaev1999quantum,aharonov2002quantum}.
The similar idea is also used to prepare the ground state of the many-body quantum systems \cite{poulin2009preparing,Nagaj2009fast}. 
Here, we study the subset-sum problems by using a quantum heuristic approach similar to the verification circuits of quantum Arthur-Merlin games\cite{marriott2005quantum}. 
The approach may provide exponential speed-up for the solution of the subset problems over the classical algorithms under the following assumptions:
	\begin{as}
		\label{assumption1}
	Let	$|L|$ be the number of possible subset-sums less than $W$ and  $|L'|$ be the number of possible subset-sums greater than or equal to $W$. In this paper, we will assume that $\frac{|L'|}{|L|} = poly(n)$. 
	\end{as}
	Under this assumption, the probability of having a subset-sum less than $W$ is $\frac{1}{poly(n)}$. However, we may still have $O\left(\frac{2^n}{poly(n)}\right)$ number of possible subsets which gives a sum less than $W$. 
	Whence we can easily make the following remark:
	\begin{rem}
		The subset-sum problem under this assumption is as difficult as without this assumption. Therefore, the maximization version of the problem is still NP-hard when $W = O(2^n)$. 
	\end{rem}
In addition, the correctness of the solution produced by our heuristic is determined from the distribution (which can be guessed from the distribution of the input elements) of the feasible subset-sums. 
	It yields the exact answer with a high-probability if the following condition is satisfied.
		\begin{as} 
			\label{assumption2}
		Let $\phi_{max}$ be the maximum subset-sum less than $W$. 
		Let $m$-qubits in the output register of our algorithm represents the binary value of $\phi_{max}$.  
			If the bit value of any $t$th qubit is 1 in the binary value of $\phi_{max} = (b_0\dots b_{m-1})_2$; then after measuring the first $(t-1)$ number of qubits with the correct values $(b_0b_1\dots b_{t-2})_2$, the probability of seeing \ket{1} on the $t$th qubit is not exponentially small (i.e. the probability is $1/poly(n)$.) in the normalized-collapsed state.
		\end{as}
This assumption (condition)  only affects the accuracy of the output. As we shall show in the following sections, it does not change the polynomial running time. 
In addition, to the best of our knowledge, this assumption  does not simplify the original problem for the current classical algorithms:
 For instance, this condition is likely to hold when we have a random uniformly distributed set of input elements. 
As mentioned, generating possible subset-sums alone takes exponential time for any classical algorithm and the computational complexity of finding the solution is still bounded by $O(2^n)$ for any classical algorithm when $W=O(2^n)$.

	In the following sections, after preliminaries, we list the algorithmic steps and explain each step in the subsections. Then, we discuss the complexity analysis and show how the approach takes $O(poly(n))$ time under the above assumptions. We also discuss the application to the knapsack problem. Finally, we present a numerical example and conclude the paper.

\section{Preliminaries}
	In this section, brief descriptions of the quantum algorithms used in this paper are given. For a broader understanding of these algorithms, the reader should refer to the introductory book by Chuang and Nielsen \cite{nielsen2002quantum}.
	\subsection{Notes on Notations}
	Throughout the paper, we will use \ket{...} to represent a quantum state (a vector) and \bra{...} to conjugate transpose of a vector. 
	Bold faces such as \ket{\bf0} indicates the vector is at least a two-dimensional vector. Using a value inside the ket-notation such as \ket{\phi} indicates the basis vector associated with the binary representation of $\phi$. 
	Other than $W$ which is a given value, the capital letters generally represent matrices (operators). The quantum state \ket{1} on a qubit represents 1 as a bit value and \ket{0} is 0. The indices start from 0.
	
	\subsection{Quantum Phase Estimation Algorithm}
	Quantum phase estimation algorithm (PEA)\cite{kitaev1995quantum} is a well-known eigenvalue solver which estimates the phases of the eigenvalues of a unitary matrix, $U\in C^{\otimes n}$: i.e. the eigenvalues of $U$ comes in the form of $e^{i\phi 2\pi}$ with an associated eigenvector $\ket{\psi}$. The  algorithm estimates the value of $\phi$ for a given approximate eigenvector $\ket{\psi}$. 
	The accuracy of the estimation is determined by the overlap of the approximate and actual eigenvectors and the number of qubits used to represent the phase value. 
	PEA in general requires two registers to hold the value of the phase and the eigenvector. 
	The algorithm starts with an initial approximation of the eigenvector on the second register and \ket{\bf{0}} state on the first register: \ket{\bf{0}}\ket{\psi}.
	Then, the quantum Fourier transform is applied to the first register. It then the controlled-operators $U^{2^j}$s are applied to the second register in consecutive order: here, $0\leq j< m$, each $U^{2^j}$ is controlled by the $j$th qubit of the first register and  $m$ is the size of this register which determines the decimal precision of the estimation. 
	At this point in the first register the Fourier transform of the phase is obtained. Therefore,  applying the inverse Fourier transform and measurement on the first register yields the estimation of $\phi$. 

	In general,	the computational complexity of PEA is governed by the number of gates used to implement each $U^{2^j}$. When they can be implemented in polynomial time, then the complexity of the algorithm can be bounded by some polynomial time, $O(poly(n))$.
	\subsection{Amplitude Amplification Algorithm}
	The amplitude amplification (AA) is based on the Grover search algorithm \cite{grover1997quantum} and used to amplify the part of a quantum state  which is considered as ``good". The algorithm is mainly composed of two operators. The first operator, $F$, marks (negates the signs of) the ``good" states and the second operator, $S$, amplifies the amplitudes of the marked states.
	
	For $\ket{\psi} = \alpha \ket{\psi_{good}} + \beta \ket{\psi_{bad}}$; 
	$F\ket{\psi} = \alpha \ket{\psi_{good}} - \beta \ket{\psi_{bad}}$. The implementation of $F$ depends on the function that describes the good part of the states. Through this paper, $F$ is simply some combination of controlled $Z$ and $X$ gates ($X$ and $Z$ are Pauli spin matrices.).
	
	If $\ket{\psi} = A \ket{\bf{0}}$ for some  unitary matrix $A \in C^{N}$, then $S = 2\ket{\psi}\bra{\psi} - I = A U_{0^\perp}A^*$, where ``*" represents the conjugate transpose. 
	The amplification is done by applying the iterator $G = SF$ consecutively to \ket{\psi}. The number of iteration depends on $\beta$ and is bounded by $O(\frac{1}{\beta})$. For further details and  variants of AA, please refer to Chapter 8 of Ref.\cite{kaye2007introduction}.

\section{Algorithm}
	The approach uses a qubit and a rotation phase gate for each element of $V$,   to encode the possible subset sums as the eigen-phases of a diagonal unitary matrix $U$. 
	Then, it applies the phase estimation algorithm  to obtain the possible sums and associate eigenvectors on two quantum registers. 
	Marking the states with phases less than $W$, it eliminates the states with phases greater $W$ through the amplitude amplification.
	Finally, it again employs the amplitude amplification in measurement processes to obtain the maximum phase and its associated eigenvector which indicates the solution of the problem.
Here, first the algorithmic steps in general are listed, and then the explanations and more details for each steps are given in the following subsections.
The steps are generalized as follows:
	\begin{enumerate}
		\item Encode the integer values as the phases of the rotation gates aligned on different qubits.  
		\item Apply the phase estimation algorithm to the equal superposition state so as to produce the phases and the associated eigenvectors on quantum registers.
		\item Apply the amplitude amplification to eliminate the states where $\phi_j \geq W$.   Now, the superposition of the sums and the eigenvectors are obtained with equal probabilities.
		\item Find the maximum $\phi_j$ in the first register. Then, measure the second register to attain the solution. 
	\end{enumerate}
	The quantum circuit representing the above steps is drawn in Fig.\ref{FigCircuit1General}. 
	\subsection{Encoding the Values into the Phases}
	First, the values are scaled so that $\sum_{j =0}^{n-1}  v_j \leq 0.5$. 
	For each value $v_j$ with $0\leq j<n$, a rotation gate in the following form is put on the $(j+1)$st qubit:
	\begin{equation}
	\label{EqPhaseGate}
		R_j = \left( \begin{matrix}
		1&0\\
		0&e^{iv_j2\pi} 
		\end{matrix}\right).
	\end{equation}
  The $n$-qubit circuit formed with these rotation gates can be then represented by the following unitary matrix:
	\begin{equation}
		U  = R_{n-1}\otimes \dots \otimes R_0. 
	\end{equation} 
	Here, $U$ is a diagonal matrix with the diagonal elements (eigenvalues):
	\begin{equation}
	\left[ 1, e^{iv_0}, e^{iv_1}, e^{iv_0+v_1}, \dots, e^{i(v_0+ \dots + v_{n-1})} \right] = \left[ e^{i\phi_0}, e^{i\phi_1},  \dots, e^{i \phi_{n-1}} \right].
	\end{equation} 
	As seen in the above, the phases of the eigenvalues associated with the eigenvectors forming the standard-basis-set encode all the possible sum and the subset information: i.e. the $j$ vector in the standard basis indicates the phase $\phi_j$ and the elements of the $j$th subset.
	\subsection{Generating All Possible Sums and Subsets on Registers}
Consider the phase estimation algorithm applied to $U$ with the following initial state:
	\begin{equation}
	\ket{\psi_0} = \ket{\bf0}\frac{1}{\sqrt{2^{n}}}\sum_{j=0}^{2^n-1}\ket{j}.
	\end{equation}
Here, $\ket{\psi_0}$ can be simply generated by 
$( I^{\otimes m}\otimes H^{\otimes n}) \ket{\bf0}\ket{\bf0}$, where $I$ represents an identity matrix and $H$ is the Hadamard matrix. 
Since the eigenvectors of $U$ are of the standard basis, the final output of the phase estimation holds the equal superposition of the eigenvector and phases:
	\begin{equation}
	\label{EqAfterPEA}
	\ket{\psi_1} = U_{pea}\ket{\psi_0} = \frac{1}{\sqrt{2^{n}}}\sum_{j=0}^{2^n-1}\ket{\phi_j}\ket{j},
	\end{equation} 
where $U_{pea}$ represents the phase estimation algorithm applied to $U$ and  forms the first part of the circuit in Fig.\ref{FigCircuit1General}.
Obviously if we are able to efficiently find the index of the maximum $\phi_j$ less than  $W$ in the above, then we solve the maximum subset-sum problem efficiently.

\subsection{Eliminating the Subsets with $\phi_j\geq W$}
Before searching for the solution, we divide the quantum state in Eq.\eqref{EqAfterPEA} into two parts: 
\begin{equation}
\label{EqAfterPEA2}
\ket{\psi_1} = 
 \frac{1}{\sqrt{2^{n}}}\sum_{j\in L}\ket{\phi_j}\ket{j} + \frac{1}{\sqrt{2^{n}}}\sum_{j\in L'}\ket{\phi_j}\ket{j} .
= \sqrt{\frac{|L|}{2^n}} \ket{\psi_{good}} + \sqrt{\frac{|L'|}{2^n}} \ket{\psi_{bad}}
\end{equation} 
where $L =\{j:\phi_j<W\}$ and 
$L' = \{j:\phi_j\geq W\}$ with $0\leq j<2^n$.
This equation includes all the possible eigenpairs. To eliminate the ones included in $L'$, we apply the amplitude amplification algorithm defined by the iterator $G =S\left(F_{\phi} \otimes I^{\otimes n}\right)$ as shown in Fig.\ref{FigCircuit1General}. Here, $F_{\phi}$ operates on the first register and flips the sign of the states with $\phi_j  < W$: 
\begin{equation}
\left(F_{\phi} \otimes I^{\otimes n}\right)\ket{\psi_1} = -\sqrt{\frac{|L|}{2^n}}\ket{\psi_{good}} + 
\sqrt{\frac{|L'|}{2^n}} \ket{\psi_{bad}}. 
 \end{equation}
$S = 2\ket{\psi_1}\bra{\psi_1}-I$ and can be implemented as follows:
	\begin{equation}
	\label{EqS}
	S = (I^{\otimes m}\otimes H ^{\otimes n})U_{pea} U_{0^\perp}(I^{\otimes m}\otimes H ^{\otimes n})U_{pea}^*,
	\end{equation}
where  $U_{0^\perp} = I - 2\ket{\bf0}\bra{\bf0}$.

After each iteration, the amplitudes of the ``good" states are amplified.
The number of iterations in the algorithm (the number of applications of $G$) is determined by the initial probability and  is bounded by $O\left( \sqrt{\frac{2^n}{|L|}}
\right)$. In the worst case where $|L|<<|L'|$, the complexity becomes $O(\sqrt{2^n})$. 
However, in the other cases, the number of iterations is bounded by $O(poly(n))$. In addition, using the quantum counting one can estimate the value of $\sqrt{\frac{|L|}{2^n}}$ and $|L|$ in polynomial time (see quantum counting in Chapter 8 of Ref.\cite{kaye2007introduction}). 

As explained in the Introduction, in this paper we make the assumption given in Assumption \ref{assumption1}: i.e., mainly $\frac{|L'|}{|L|} = O(poly(n))$. 
	As a result of this assumption, this part of the algorithm takes $O(poly(n))$ time. And at the end of the amplitude amplification, the final quantum state  becomes:
	\begin{equation}
	\label{Eqpsi2}
	\ket{\psi_2} \approx \frac{1}{\sqrt{|L|}}\sum_{j \in L}\ket{\phi_j}\ket{j}.
	\end{equation}

\subsection{Finding the Maximum Sum with Its Subset}
	Grover search algorithm \cite{grover1997quantum} is able to find a maximum or minimum element of a list of $|L|$ items in $O(\sqrt{|L|})$ times \cite{durr1996quantum}. 
	It can be directly applied to Eq.\eqref{Eqpsi2} to find the maximum of $\phi_j$s and the value of $j$. However, this makes the running time of the whole algorithm exponential because of Assumption \ref{assumption1}. 
	
	The elements of the set $\{\phi_j:0 \leq j<2^n\}$ is partially sorted and mostly $\phi_j \leq \phi_{j+x}$ for a considerably large $x$. Therefore, in some cases, quantum binary search algorithm (e.g. \cite{ambainis1999better,hoyer2002quantum}) can be used to produce an approximate solution. This will require $O(lg{|L|}) = O(poly(n))$ time complexity.
	
	Below, similarly to the binary search algorithm and the verification circuit given \cite{marriott2005quantum}, a polynomial time method for finding the maximum is presented by applying a sequence of conditional amplitude amplifications:	
	Let us assume the maximum $\phi_j$ in Eq.\eqref{Eqpsi2} is $\phi_{max} = (b_0\dots b_{m-1})_2$. If we try to maximize the measurement outcome of the first register, then we attain a value close to $\phi_{max}$. 
	This maximization can be done by starting the measurement from the most significant qubits while trying to measure  as many qubits in \ket{1} state as possible. Therefore, we will measure a qubit: if the outcome is not \ket{1}, then we apply the amplitude amplification to amplify the states where this qubit is in \ket{0} state and then do the measurement again. If the qubit does not yield \ket{1} after a few iterations, then we will assume \ket{0} as the qubit value and move on to the next qubit. And we repeat this process for the all qubits in the first register. This is explained in more details below and indicated in Fig.\ref{FigCircuit1General} (Note that the measurements after each amplitude amplification is omitted in the figure.): 
	
	\begin{itemize}
		\item We measure the first qubit (representing the most significant bit, $b_0$): 
		\begin{itemize}
			\item If it is \ket{1}, then we set \ket{b_0} = \ket{1} and move on to the next qubit in the collapsed state. 
			\item Otherwise, we apply the amplitude amplification by flipping the signs of the states in which the first qubit is in \ket{1} state. While the flipping can be done by simply using the Pauli $Z$-gate, the amplification operator,  $S_1$, can be implemented in a way similar to Eq.\eqref{EqS}:
				\begin{equation}
				\label{EqS1}
				S1 = U_1 U_{0^\perp}U_{1}^*,
				\end{equation}
   where $U_1$ represents all the quantum operations up to this point. Here, we only apply the  iterator $G_1 = S_1 \left(Z\otimes I^{\otimes mn-1}\right)$  a few times until the measurement yields \ket{1}. If it does not, then \ket{b_0} is set to \ket{0}.
		\end{itemize} 
		\item   In the second qubit, we repeat the same process. However, this time 
		$G_2 = S_2 \left(\hat{Z}\otimes I^{\otimes mn-2}\right)$; where, $S_2$ involves all the operations done up to this point, and the gate $\hat{Z}$ is the controlled-$Z$ gate acting on the second qubit and controlled by the first qubit: if the first qubit was \ket{0}, then the control bit is \ket{0} (i.e. the gate acts when the first qubit is \ket{0}.). Otherwise, it is set to \ket{1}. 
		\item Similarly, using $Z$ gates controlled by the previous qubits, the measurements along with the amplitude amplifications are repeated for the remaining qubits. Here, the control-bits are either 0 or 1 determined from the measurement results of the previous qubits.
	\end{itemize}
	The above maximization method is able to amplify the amplitude of $\phi_{max}$ if at any point the probability to measure $\ket{1}$ on the qubit is not exponentially small. Otherwise, the number of the amplitude amplification required to see \ket{1} on that qubit becomes exponentially large. 
	Since the amplitude amplification  is only applied a few times (when \ket{1} is not encountered, the qubit is assumed to be \ket{0}), this will cause an error in the result.  
Let us assume that the condition given in Assumption \ref{assumption2} holds: i.e., if the bit value of any $t$th qubit is 1 in the binary value of $\phi_{max} = (b_0\dots b_{m-1})_2$; then after the individual measurements of the first $(t-1)$ number of qubits with the values $(b_0b_1\dots b_{t-2})_2$, the probability of seeing \ket{1} on the $t$th qubit is not exponentially small in the normalized-collapsed state.
	This assumption  affects the accuracy of the result rather than the running time since the amplitude amplification on a qubit is applied only a few times and when \ket{1} is not encountered, the related bit value of $\phi_{max}$ is assumed 0. 
	
\begin{figure*}
		\centering
		\includegraphics[width=5in]{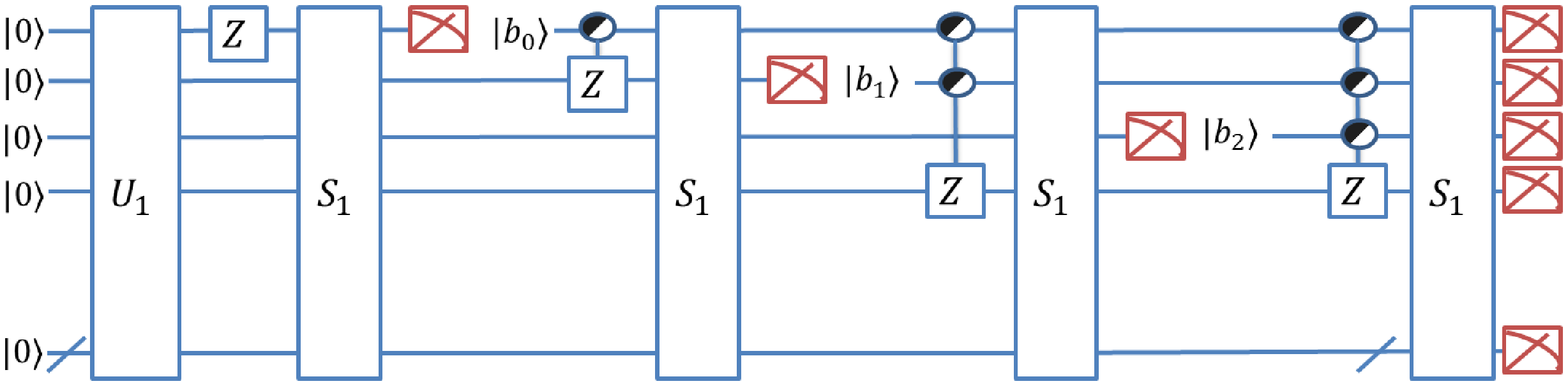}
		\caption{The circuit for the algorithm according to Remark \ref{remark1} with 4 qubits on the first register: $U_1 = \left(I^{\otimes 4}\otimes H^{\otimes n}\right)U_{pea}\left(F_\phi \otimes I^{\otimes n}\right)S$ and $S_1 = U_1U_{0^\perp}U_1^*$.
			\label{FigCircuit2General} 
		}
	\end{figure*}

	To further simplify the circuit in Fig.\ref{FigCircuit1General} and the numerical simulations, we will also make following remark:
	\begin{figure*}
		\centering
		\includegraphics[width=5in]{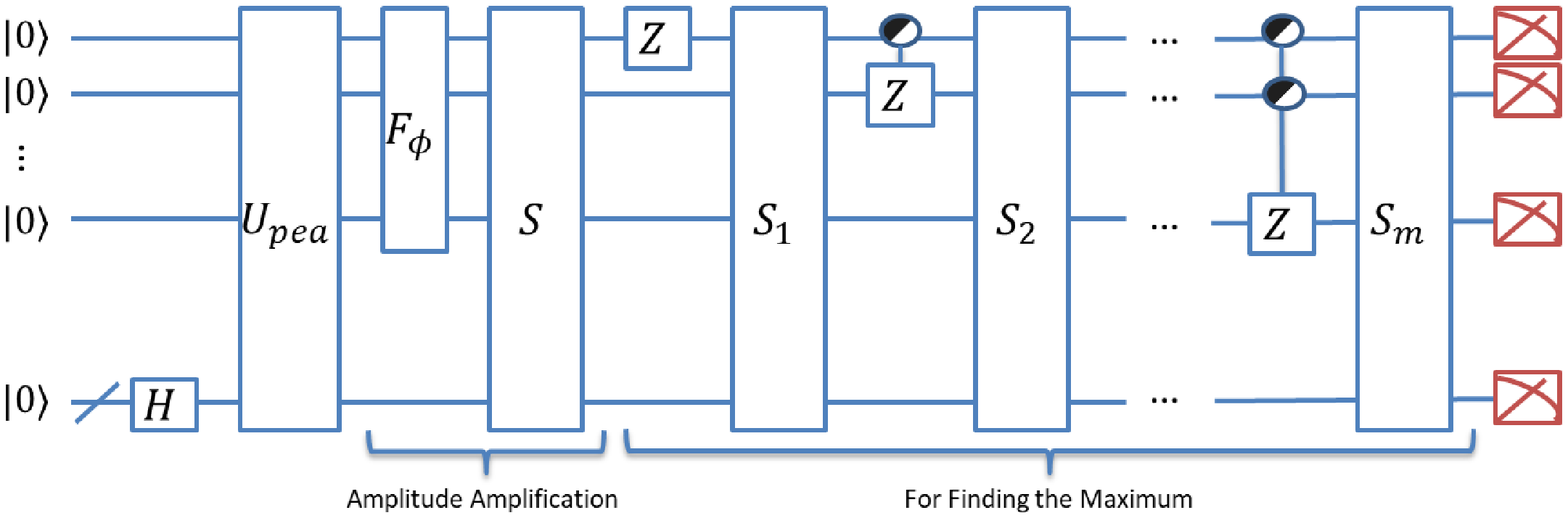}
		\caption{The general circuit for the algorithm: $S = \hat{U}_{pea}U_{0^\perp}\hat{U}_{pea}^*$ with $\hat{U}_{pea} = \left(I^{\otimes m}\otimes H^{\otimes n}\right)U_{pea}$. And $S1 = \hat{U}_{pea}\left(F_\phi \otimes I^{\otimes n}\right)SU_{0^\perp}S^*\left(F_\phi^* \otimes I^{\otimes n}\right)\hat{U}_{pea}^*$ and so on.
			\label{FigCircuit1General} 
		}
	\end{figure*}
\begin{rem}
		\label{remark1}
		 $S_1$ can be used in places of $S_2\dots S_m$ to simplify the implementation of the amplitude amplifications.
\end{rem}
	The circuit in accordance with the above remark is presented in Fig.\ref{FigCircuit2General} (The measurements on the qubits are also explicitly indicated in this figure however not in Fig.\ref{FigCircuit1General}).  Sect. VI gives a numerical example based on this circuit.
	Now, we will explain  how this circuit may yield the solution by going through the measurements of the first two qubits on the circuit:
Let us first divide the state given in Eq.\eqref{Eqpsi2} (the state after $U_1$ in Fig.\ref{FigCircuit2General}) into four parts with the same length:
		\begin{equation}
		\ket{\psi_2} = 
		\left(
		\begin{matrix} \bf{x_0}\\
		\bf{x_1}\\
		\bf{x_2}\\
		\bf{x_3} 
		\end{matrix}
		\right).
		\end{equation}
	The probabilities of measuring \ket{0} and \ket{1} on the first qubit are $P_0 = (||\bf{x_0}||^2 + ||\bf{x_1}||^2 )$ and $P_1 = (||\bf{x_2}||^2 + ||\bf{x_3}||^2 )$, respectively.
	If $P_1$ is not exponentially less than $P_0$, then with the help of the amplitude amplification ($G_1 = S_1\hat{Z}$), the first qubit, can be measured in \ket{1}. Therefore, $b_0$ becomes 1.
    Let us assume we obtain $b_0=1$ after the measurement. 
		If we use a qubit  in place of the first qubit and initialize it in \ket{1} state, we obtain the following normalized-sate:
		\begin{equation}
		\ket{\psi_3} =
		\ket{1} \otimes \left( 
		\begin{matrix}  \zeta \bf{x_2} \\ \zeta \bf{x_3}\end{matrix}\right),
		\text{\ with\ } \zeta = \frac{1}{\sqrt{||\bf{x_2}||^2+||\bf{x_3}||^2}}.
		\end{equation}
		For the second qubit,  the part represented by \bf{x_3} is marked by the controlled $Z$ gate, and then $S_1$ is applied:
		\begin{equation}
		S_1(\hat{Z}\otimes I)\ket{\psi_3}
		=\left(2\ket{\psi_2}\bra{\psi_2}-I\right)
		\left( 
		\begin{matrix} \bf{0}\\ \bf{0}\\ \zeta\bf{x_2} \\ -\zeta\bf{x_3}\end{matrix}\right)
		=\zeta \left(\begin{matrix} 
		2d_x \bf{x_0}\\
		2d_x\bf{x_1}\\
		(2d_x-1)\bf{x_2}\\
		(2d_x+1) \bf{x_3} 
		\end{matrix}\right),
		\end{equation}
		where $d_x = ||\bf{x_2}||^2- ||\bf{x_3}||^2$. 
		Due to $||\bf{x_2}|| \geq ||\bf{x_3}||$, $d_x\geq 0$:
		\begin{itemize}
			\item If $dx \geq 0.5$, then all of the amplitudes in the above quantum states are unmarked. Therefore, in the subsequent iteration of AA only $\bf{x_3}$ will  be marked, and the amplitudes corresponding to $\bf{x_3}$ will be amplified.
			\item If $dx = (||\bf{x_2}||- ||\bf{x_3}|| )< 0.5$, that means the probability difference between \ket{0} and \ket{1} is small; thus, we are very likely to measure \ket{1}  after the first amplitude amplification. After the measurement, the state collapses to 
			\begin{equation}
			\zeta \left(\begin{matrix} 
			\bf{0}\\
			2d_x\bf{x_1}\\
			\bf{0}\\
			(2d_x+1) \bf{x_3} 
			\end{matrix}\right).
			\end{equation}
		\end{itemize}
	
\section{Complexity Analysis}
 We will follow the circuit in Fig.\ref{FigCircuit1General} to analyze the complexity of the whole approach under Assumption 1 and 2. The algorithm starts with two quantum registers of respectively $m$ and $n$ qubits, and then later in the maximum finding part of the algorithm it uses another register with $m$ qubits which is implicitly indicated in Fig.\ref{FigCircuit2General} but not  in Fig.\ref{FigCircuit1General}. Therefore, the total number of qubits employed in the whole running of the algorithm is $(2m+n)$.

Since $U$ involves only phase gates described in Eq.\eqref{EqPhaseGate}, using $U^{2^j}$, $0 \leq j<m$ in the phase estimation requires only $n$ number of controlled phase gates (Note that the power of the unitary can be taken by simply changing the angles of the rotation gates.).
	Therefore, including the complexity of the quantum Fourier transform \cite{nielsen2002quantum}, $U_{pea}$, the phase estimation part, requires $O(n+mlgm) =O(poly(n))$ number of quantum gates. 
	
	In the amplitude amplification part, the operator $F_{\phi}$ can be  designed in $O(poly(n))$ time by using a logical circuit: i.e., the circuit is composed of $X$ and $Z$ gates and marks all of the states less than ${W}$.   
Moreover, the implementation of $S$ defined in Eq.\eqref{EqS} involves the Hadamard gates, $U_{pea}$ and $U_{0^\perp}$ all of which  can be implemented in polynomial time. 
	
The remaining part of the circuit is for finding maximum and involves  $S_j$s and controlled-$Z$ gates. The implementation of any $S_j$ is similar to the operator $S$: they involve the repetitions of the circuit up to their location, and hence requires more computations. However, because of Assumption 2, this part of the circuit and the whole processes are still bounded by some polynomial time, $O(poly(n))$.

\section{Application to 1/0-Knapsack Problem}
The maximum subset-sum problem is related to many other problems. One of these  is the 1/0-knapsack problem \cite{kellerer2004introduction} described as: For a given items with weights $\{w_0 \dots w_{n-1}\}$ and values $\{v_0, v_{n-1}\}$, determine which items should be included in a subset to maximize the total subset-value while keeping the total-weight less than $W$.
	This problem can be solved in a similar fashion to the subset-problem by adding one additional register to the algorithm:
	\begin{itemize}
		\item \ket{sw} -- The first register holds the sum of the weights.
		\item \ket{sv} -- The second register holds the values.
		\item \ket{j} -- The third register indicates the items included in the subset: \ket{j} describes the $j$th vector in the standard basis.
	\end{itemize}
	The algorithm starts with constructing the superposition of the possible sums of weights and values as a quantum state by using the phase gates where the least-significant-bits encode the item-values while the most-significant bits are used for the weights. After the phase estimation, the following quantum state is generated:
	\begin{equation}
	\ket{\psi_1} = 
	\frac{1}{\sqrt{N}}\sum_{j=1}^N\ket{sw_j}\ket{sv_j}\ket{j}
	\end{equation}
	Then, applying the amplitude amplification, the states where $sw_j \geq W$ are eliminated and  the probability of the states where  $\ket{sw_j} < W$ are put into the equal superposition:
	\begin{equation}
	\ket{\psi_2} = 
	\frac{1}{\sqrt{|L|}}\sum_{j\in L}\ket{sw_j}\ket{sv_j}\ket{j}
	\end{equation}
	Now, the maximum finding is done on the second register:  After finding \ket{sv_j} with the maximum decimal value, the solution to the knapsack problem 
	is obtained from the corresponding \ket{j} which indicates the involved items. 

	\section{Numerical Example}
	In this section, we present a random numerical example based on the circuit in Fig.\ref{FigCircuit2General}.
	Let us assume that given the set of values 
	\begin{equation}
	V = \left\{
	\begin{matrix}
	0.10937500,\\
	0.10546875,\\
	0.10156250,\\
	0.09375000,\\
	0.05468750,\\
	0.02343750,\\
	0.00390625
	\end{matrix}  \right\},
	\end{equation} 
	which are normalized so that the maximum possible sum is at most 0.5, we are asked to find the subset which gives the maximum possible sum \underline{less than} $W = 0.19921875$. If we use 9 bits of precision: i.e. $m=9$ (the size of the first register in PEA), then $W=(001100101)_2$).   
	
\begin{figure}
		\centering
		\includegraphics[width=3in]{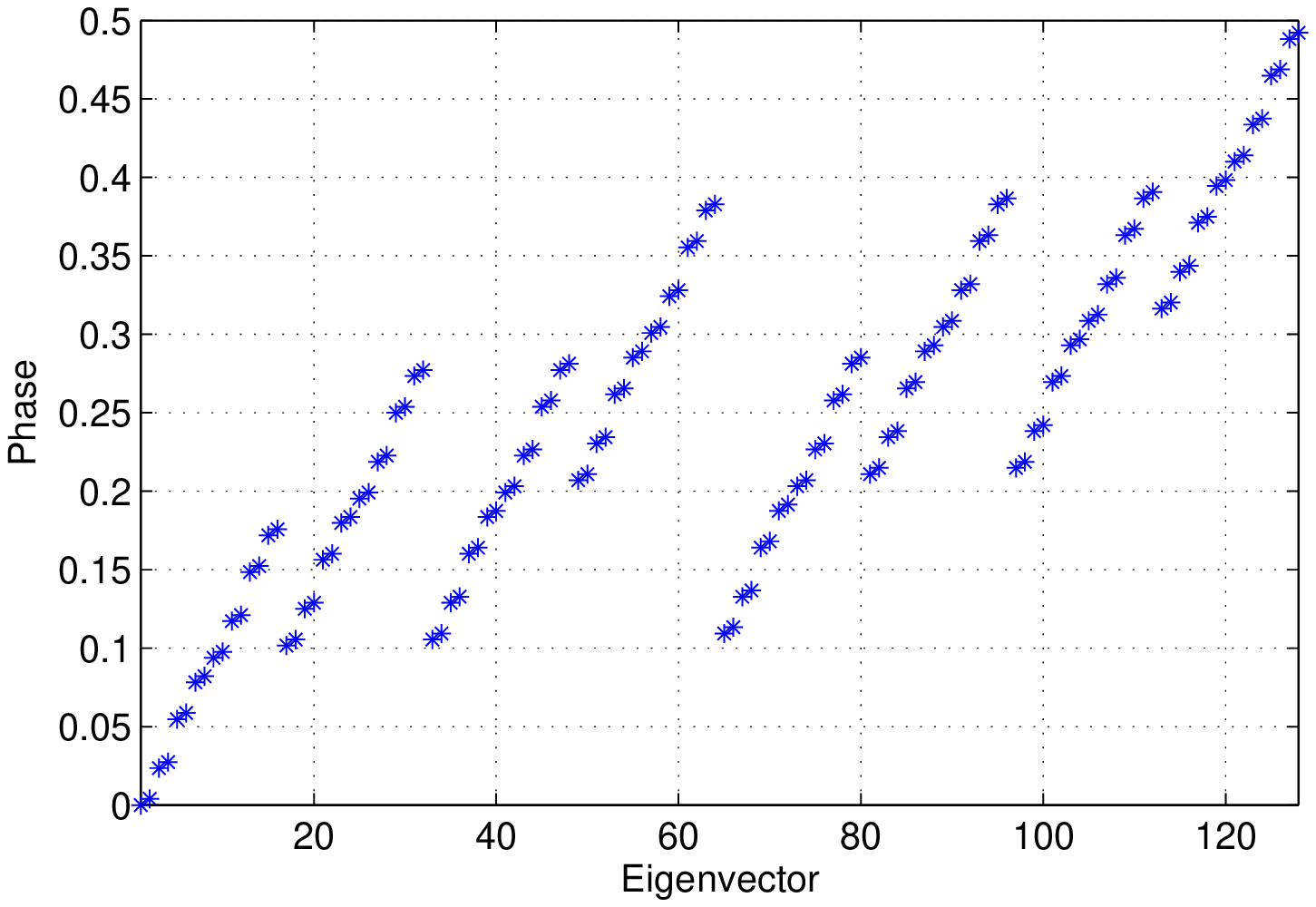}
		\caption{Phases.
			\label{FigEigenvalues}}
	\end{figure}

\begin{figure}
		\centering
		\includegraphics[width=3in]{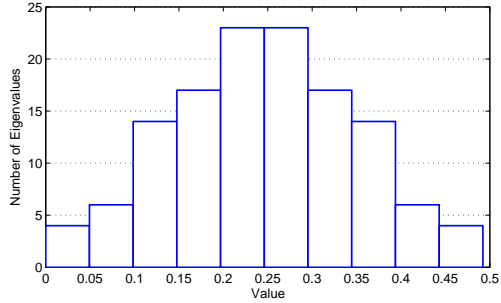}
		\caption{Distribution of the eigen-phases.
			\label{FigHist} 
		}
	\end{figure}

	We start with the construction of $U$ which requires a qubit and a single rotation gate for each element of $V$: that means $n=7$ (the size of the second-register in PEA) and $U\in C^{\otimes n}$. 
	The eigen-phases of $U$ shown in Fig.\ref{FigEigenvalues} represents the solution space, the possible subset-sums of $V$.
 Fig.\ref{FigHist} depicts the distribution of these phases.
	
	
\begin{figure*}[h]
		\centering
		\subfloat[The probabilities of the eigenpairs after the amplitude amplification.  \label{FigProbAA}]{ \includegraphics[width=2.5in]{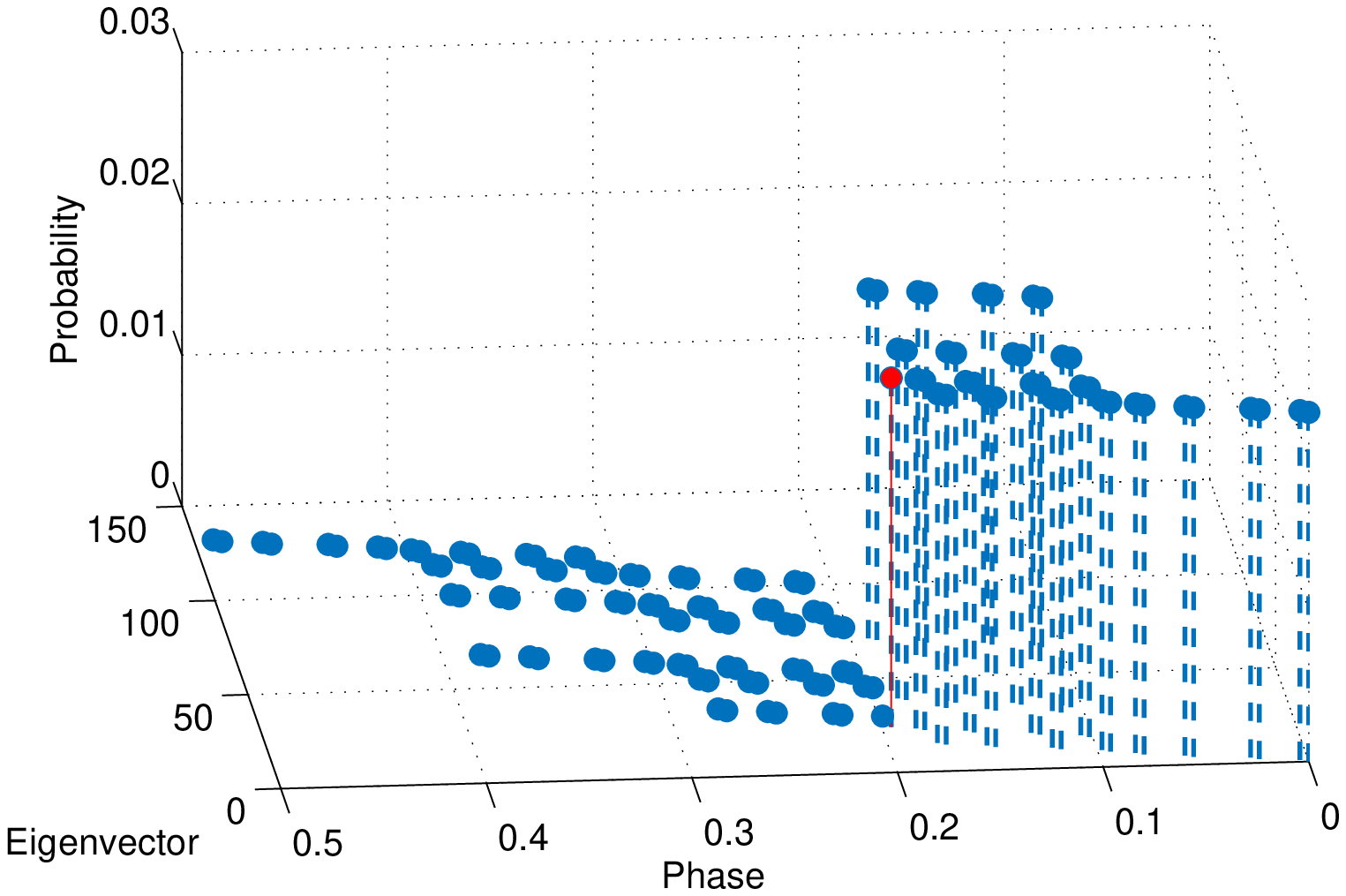}}\ 
		\subfloat[The probabilities after the measurements of qubit-1 and qubit-2 in \ket{0} states with probability 1.\label{FigProbQ1}]{ \includegraphics[width=2.5in]{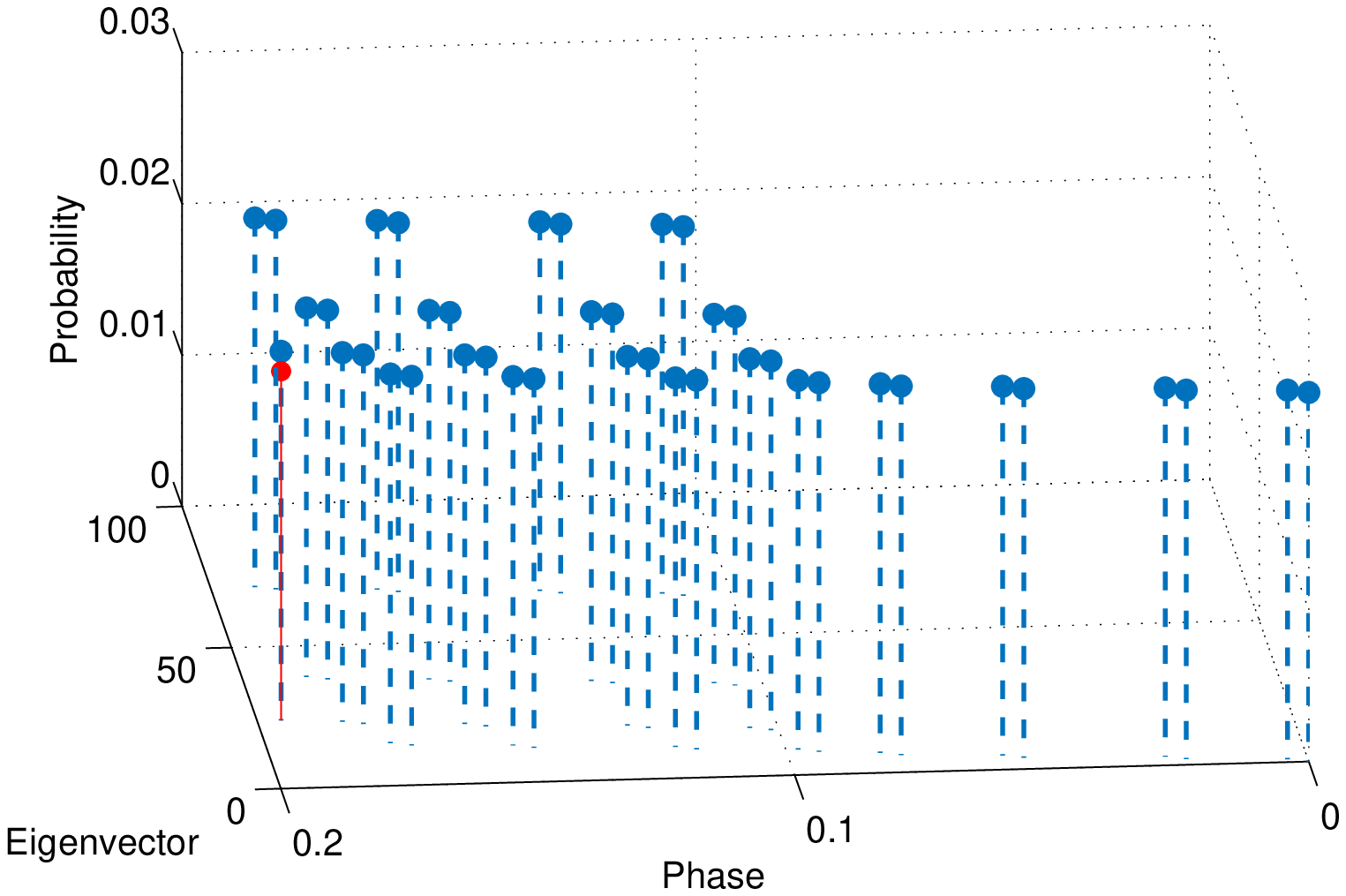}}\\
		\subfloat[The probabilites after the measurement of qubit-3 in \ket{1} state with probability 0.5610.\label{FigProbQ3} ]{ \includegraphics[width=2.5in]{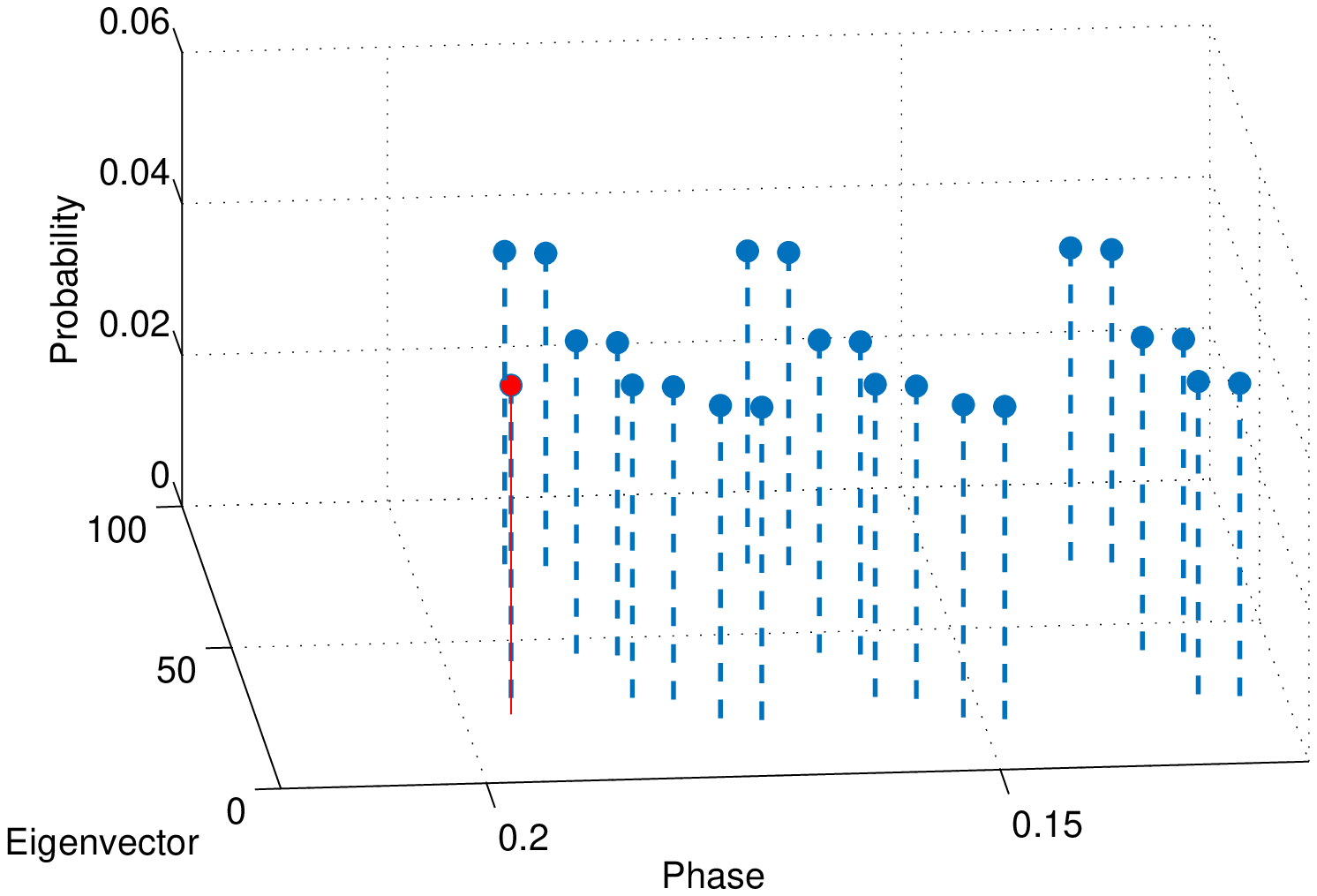}}\ 
		\subfloat[The probabilities after AA is apllied for the fourth qubit. The measurement of qubit-4 yields \ket{1} state with probability 0.8009:The probability was 0.1739 after  one iteration of AA, it is amplified to 0.8009.\label{FigProbQ4}]{ \includegraphics[width=2.5in]{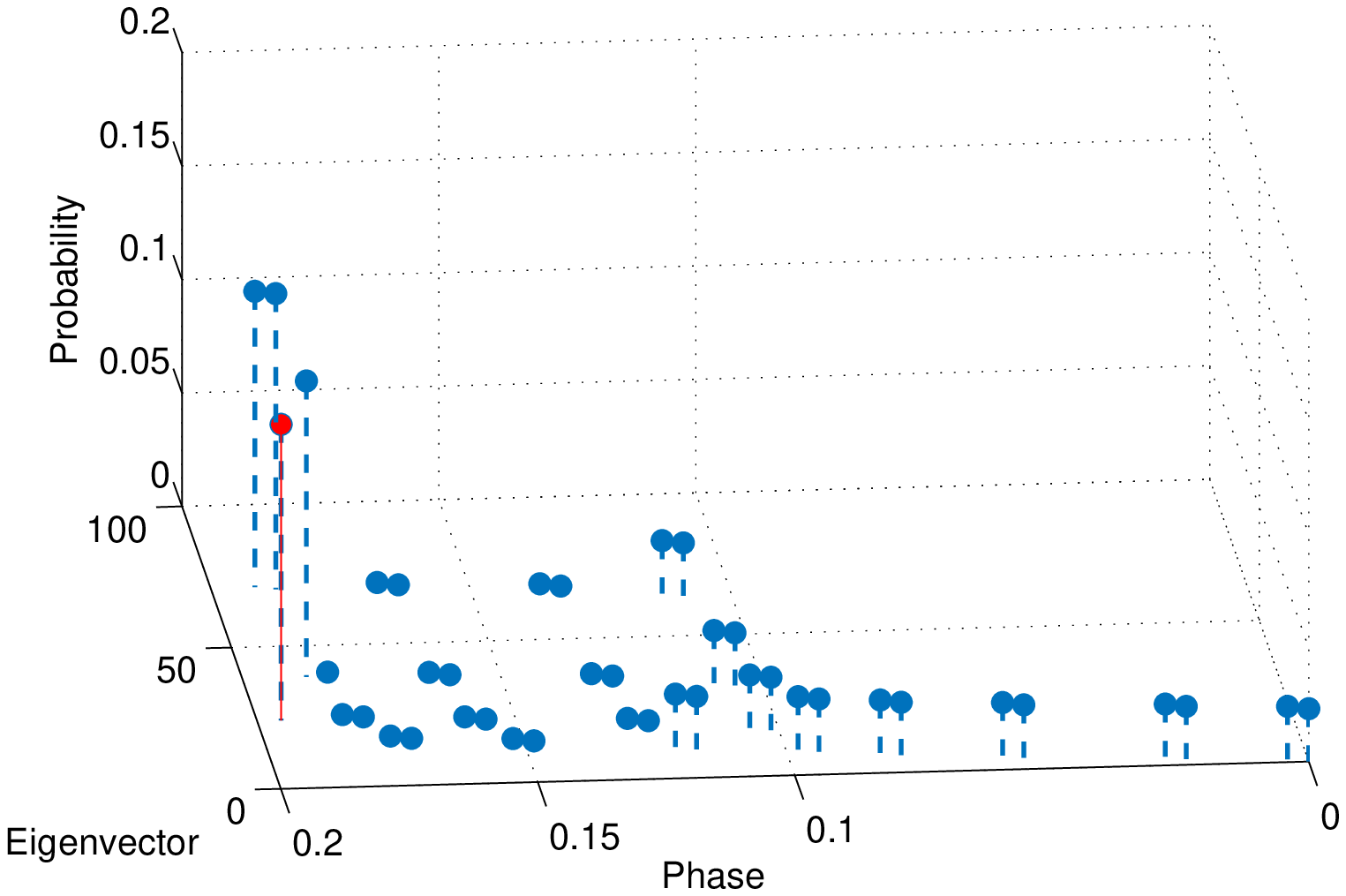}}\\
		\subfloat[The probabilities after the measurements of qubit-4 in \ket{1} and qubit-5 and qubit-6 in \ket{0} states with the probabilities 0.8009, 1, and 1, respectively.\label{FigProbQ42}]{ \includegraphics[width=2.5in]{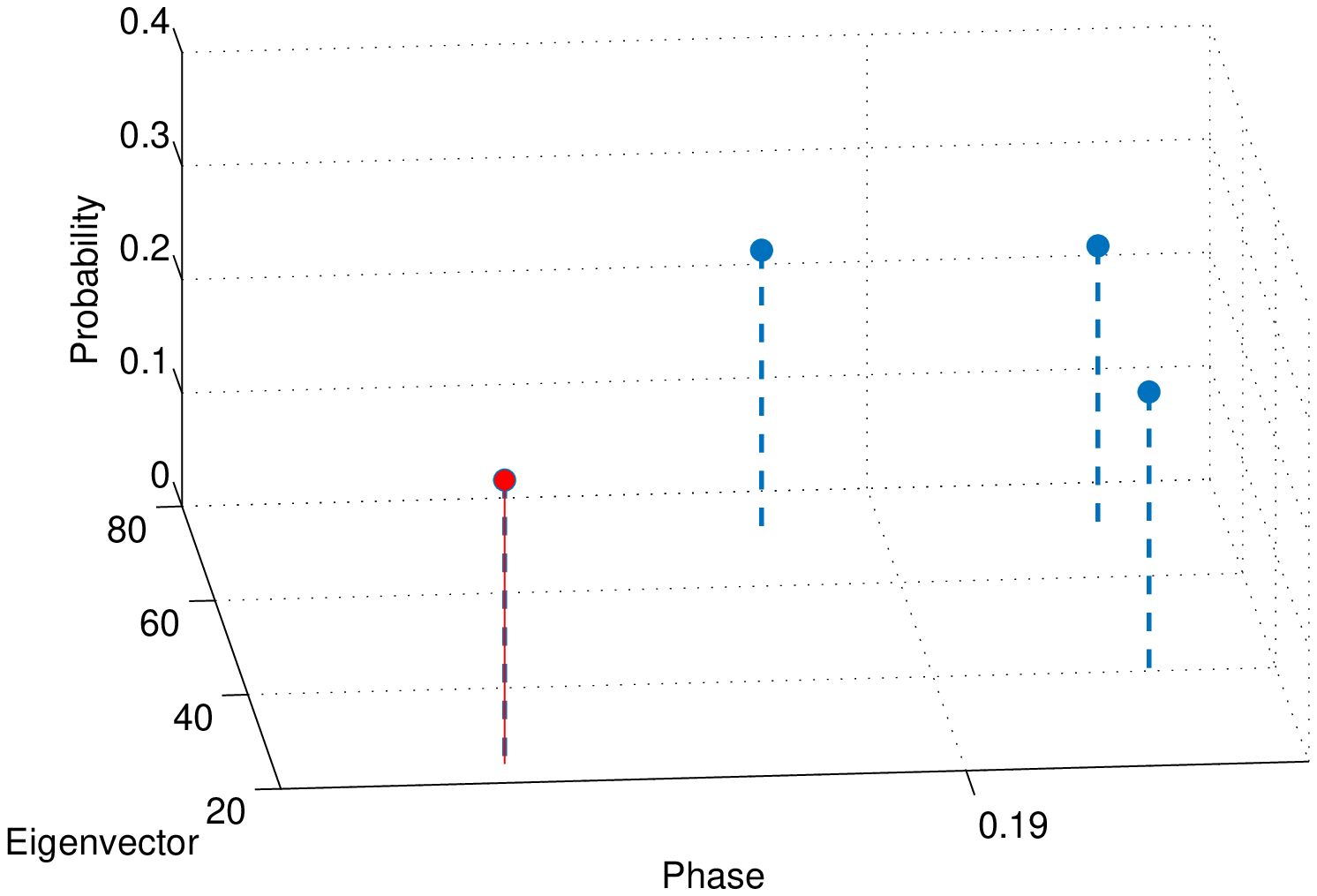}}\ 
		\subfloat[The probability of the states after AA is applied for qubit-7. The measurement on qubit-7 yields \ket{1} with probability 0.3488: The probability was 0.25, it is amplified to 0.3488 through one iteration of AA. At this point, the state collapses to the solution.\label{FigProbQ7}]{ \includegraphics[width=2.5in]{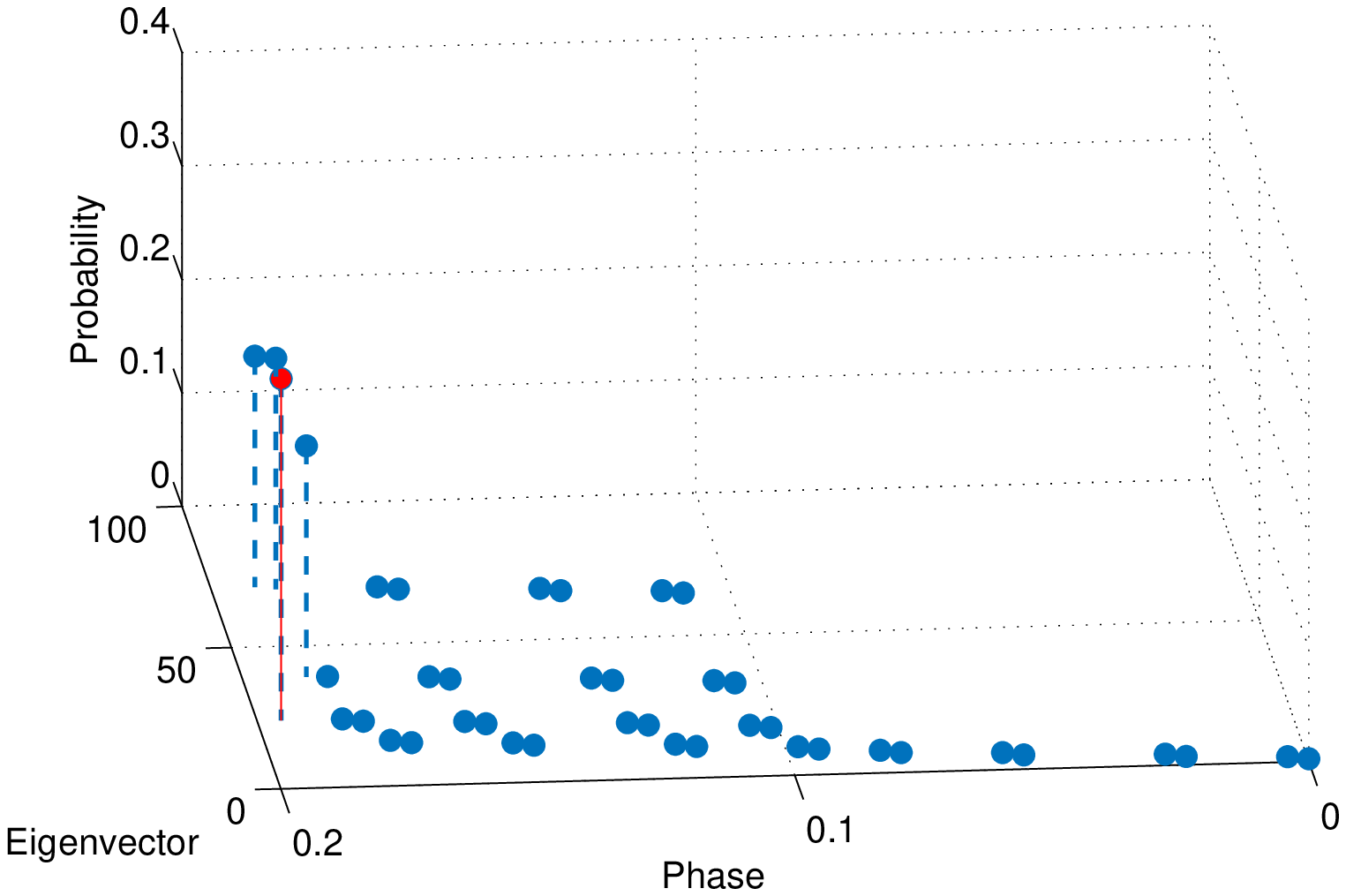}}
		\caption{The probabilities of the states after the amplitude amplification and the measurements of qubits according to the circuit given in Fig.\ref{FigCircuit2General}.}
	\end{figure*}

	After PEA is applied with an initial superposition state on the second register; using AA, the amplitudes of the states in which the phase value on the first register is less than $W=(001100101)_2$ are marked and amplified. After one iteration of AA, the probability of the eigenvector-phase pairs are presented in Fig\ref{FigProbAA}. 
	At this point, we have $\sum_{j, \phi_j<W}\ket{\phi_j} \ket{j}$.  
		
	In the maximum-finding part, we start doing the measurement from the most significant qubits: 
	\begin{itemize}
		\item  Qubit-1 and qubit-2 yields \ket{0}s with probability $\approx1$s. Therefore, \ket{b_0b_1} = \ket{00}. After the normalization, we have the solution space where all $\phi_j$s are less than $W$. This is shown in Fig.\ref{FigProbQ1}.
		\item 
		The measurement on qubit-3 yields either \ket{0} or \ket{1} with probabilities, respectively, $0.4390$ and $0.5610$. Therefore,  it is very likely \ket{1} is measured after a few attempts. 
		In that case, we have \ket{b_0}\ket{b_1}\ket{b_2}=\ket{0}\ket{0}\ket{1} on the first three qubits and the collapsed-state on the remaining qubits. 
		The normalized probabilities at this point are drawn on Fig.\ref{FigProbQ3}.
		\item 
The probabilities of \ket{0} and \ket{1} for qubit-4 is 
		$0.8261$ and $0.1739$ in the normalized state. 
		After a few measurements, if we see \ket{1}, we set \ket{b_4} = \ket{1} and continue on the fifth qubit. Otherwise, we apply the amplitude amplification. This changes the probabilities for qubit-4 to    $0.1991$ and
		$0.8009$, respectively. However, it also brings back some of the eliminated states with some small probabilities. This is shown in Fig.\ref{FigProbQ4}.
		After the measurement of qubit-4 in \ket{1}, the state collapses into four remaining eigenpairs with equal probabilities shown in Fig.\ref{FigProbQ42}.
		\item In all of these four states, qubit-5 and qubit-6 are in \ket{0} state. Therefore, the measurements on these qubits yield \ket{0} states with probability 1.
		\item  The probability of seeing qubit-7 in \ket{1} state is 0.25. An iteration of AA amplifies this to 0.3488 as shown in Fig.\ref{FigProbQ7}. At this point, after a few measurements, we are likely to encounter \ket{1} on qubit-7. This collapses the state into the solution.
		\item And finally, qubit-8 and qubit-9 are measured in \ket{0} with probability 1.	 
	\end{itemize}
	The above maximization yields 0.1953125 as the maximum phase with \ket{0011000} as the corresponding eigenvector, which is the exact solution. 
	
	\section{Conclusion}
In this paper we have studied the subset-sum and similar problems: e.g. the knapsack problem. In particular, we have generated the possible sums by using the phase estimation and the amplitude amplification algorithms. Then, we have used a maximum-finding procedure to obtain the solution.
	The approach requires polynomial time if the number of possible sums less than or equal to the given value are not exponentially smaller than the number of possible sums greater than the value. 
	In addition, it yields the exact answer  if the probability of seeing the correct bit value  on the $t$th qubit is not exponentially small in the normalized-collapsed state after the first $(t-1)$ number of most significant bit values are correctly measured. 
	The approach is general and can be further improved for the similar NP-complete problems.

		\bibliographystyle{unsrtnat}
		\bibliography{ref}

\end{document}